\documentclass[copyright,creativecommons,noncommercial]{eptcs}
\providecommand{\event}{DICE 2010} 
\usepackage{amsthm}
\usepackage[leqno]{amsmath}
\usepackage{easybmat}
\usepackage{amssymb}
\usepackage{fancyvrb}

\newtheorem{theorem}{THEOREM}[section]

\newtheorem{lemma}[theorem]{LEMMA}

\theoremstyle{definition}
\newtheorem{definition}[theorem]{Definition}

\renewcommand{\vec}{\mathaccent"017E}  
\newcommand{\deptypes}{{\mathbb{D}}}
\newcommand{\deps}{{\mathbb{F}}}
\newcommand{\unarydep}[3]{{#1\xrightarrow{\makebox[1ex][l]{$\scriptstyle #2$}} #3}}
\newcommand{\longunarydep}[3]{{#1\xrightarrow{#2} #3}}
\newcommand{\binarydep}[4]{{{#1}\atop{#2}}\rightrightarrows {{#3}\atop{#4}}}
\newcommand{\dep}[3]{{\tt#1}\stackrel{#2}{\to}{\tt#3}}  

\newcommand{\mytag}[1]{\tag{#1}}

\newcommand\Lbjk{L_{\text{BJK}}}

\newcommand{\pass}{\texttt{:=}\,}
\newcommand{\semi}{\texttt{;}}
\newcommand{\X}{\texttt{X}}
\newcommand{\Y}{\texttt{Y}}
\newcommand{\Z}{\texttt{Z}}
\newcommand{\C}{\texttt{C}}

\newcommand{\lsem}{\mbox{$\lbrack\hspace{-0.3ex}\lbrack$}}
\newcommand{\rsem}{\mbox{$\rbrack\hspace{-0.3ex}\rbrack$}}
\newcommand{\sempar}[1]{\mbox{\lsem\pgt{#1}\rsem}}

\newcommand{\nats}{\mathbb{N}}
\newcommand{\inds}{{\mathcal{I}}}
\newcommand{\pgt}[1]{{\tt #1}}

\newcommand{\bthm}{\begin{theorem}}
\newcommand{\ethm}{\end{theorem}}
\newcommand{\blem}{\begin{lemma}}
\newcommand{\elem}{\end{lemma}}
\newcommand{\bprf}{\begin{proof}}
\newcommand{\eprf}{\end{proof}}
\newcommand{\bdfn}{\begin{definition}}
\newcommand{\edfn}{\end{definition}}
\newcommand{\be}{\begin{enumerate}}
\newcommand{\ee}{\end{enumerate}}

\begin{document}

\title{On Decidable Growth-Rate Properties
of Imperative Programs
\\\ \\\large (invited talk paper)}  
\author{
   Amir M. Ben-Amram
\institute{
School of Computer Science, Tel-Aviv Academic College\\
\email{amirben@mta.ac.il}
}}
\def\titlerunning{Growth-Rate Properties
of Imperative Programs}
\def\authorrunning{Amir M. Ben-Amram}

\maketitle   

\begin{abstract}
In 2008, Ben-Amram, Jones and Kristiansen showed that
for a simple ``core" programming language---an
imperative language with bounded loops,
and arithmetics limited to addition and multiplication---it is possible
to decide precisely whether a program has certain growth-rate
properties, namely polynomial (or linear) bounds on
computed values, or on the running time.

This work emphasized the role of the
core language in mitigating the notorious undecidability of program
properties, so that one deals with decidable problems, while allowing the
application of the technique to programs in a more realistic language.
This is done by over-approximating the semantics of the concrete program.

A natural and intriguing problem was whether more elements can be added to
the core language, improving the precision of approximation, while keeping
the growth-rate properties decidable.
In particular, the method presented could not handle a command that
resets a variable to zero.
This paper shows how to handle resets.
The analysis is given in a logical style (proof rules), and the complexity
of the algorithm is shown to be PSPACE. 
The problem is shown PSPACE-complete (in contrast, without resets, the
problem was PTIME).
The analysis algorithm
evolved from the previous solution in an interesting way:
focus was shifted from proving a bound to disproving it, and
the algorithm works top-down rather than bottom-up.
\end{abstract}

\section{Introduction}

Central to the field of Implicit Computational Complexity (ICC) is
the following observation:
it is possible to restrict a programming language syntactically so that the
admitted programs will possess a certain complexity, say polynomial time,
or polynomial output size.
Since programmers want their programs to have well-behaved complexity, this
appears at first to be a useful approach. However, 
languages designed to capture a complexity class tend to be too restrictive or
 cumbersome for practical programming. The programmer
would prefer to program normally---which means that algorithms of undesirable
complexity \emph{can} be written---and
would be happy to have them detected at compile time, just as type-related errors are.
Thus, we move into the realm of \emph{static program analysis}.

Automated complexity analysis (or ``cost analysis") as a kind of static analysis
has a long history, with classic contributions including Wegbreit~\cite{Wegbreit:75},
Rosendahl~\cite{Rosendahl89}
and Le~M\'{e}tayer~\cite{ACE}.
Today it enjoys a flurry of research.

Static analysis targets program properties
that are typically uncomputable in any Turing-complete language.
Complexity properties, such as having a polynomial running time, are no
different (in fact, they are still undecidable if termination is
guaranteed). 
The common approach in static analysis is to just give up a complete solution;
usually, one takes a \emph{conservative} approach, which means that an algorithm that
has to certify programs as ``good'' may only err by rejecting a good program. In other words,
it may be sound but incomplete.

The downside of the conservative approach is that it gives up one of the hallmarks
of algorithm theory:
studying well-defined problems and developing algorithms that actually solve
them.
Besides losing the satisfaction of proving that a goal has been achieved, one
loses the ability to precisely state what has been achieved by a new algorithm, other than
by anecdotal evidence such as examples that it succeeds on.

The approach taken in this paper establishes 
a middle path. It consists of breaking the analysis of programs
in two stages: the first is \emph{abstraction}, in which the concrete
program is replaced with an abstract one,  a simplified model of the
original; the second stage is \emph{analysis}
of the abstract program.  Abstract programs are a weak model of computation 
where the properties of interest are, hopefully, decidable. Their relation to concrete
programs may be specified precisely by first assigning an \emph{approximate semantics}
to the source
program---more precisely, one which over-approximates the behaviour
of the concrete program---then translating the program to a simplified ``core" language, 
whose semantics is equal to the approximate semantics of the original.
The over-approximation ensures that the conclusions drawn for the concrete programs
are in the conservative zone. The analysis of core-language programs becomes a new,
\emph{well-defined} problem that may well be solvable.
Another benefit of the approach is that abstract (core)
programs may be rather independent of the concrete programming language,
and their analysis more widely applicable---one only needs ``front ends'' for the concrete
languages of interest.

In areas other than complexity analysis, this approach is well known.
Probably the best-known example is the abtraction of programs into
finite automata, central to software
\emph{model checking}. Closer to complexity analysis
is termination analysis. The \emph{size-change} abstraction reduces a program
to a transition system specified by
order-based constraints~\cite{leejonesbenamram01,BA:cav09}, whose
termination is decidable.

The current paper is part of a plan to apply the ``abstract and conquer''
 approach to complexity analysis,
based on previous work by Jones, Kristiansen and the author~\cite{BJK08}.
This work defined an imperative-style core language with restricted arithmetics and 
bounded loops. The loop bounds may be computed values (this is the main source
of difficulty in analysis).  It was shown that certain growth-rate questions are decidable
for this language.
Specifically, algorithms were given to decide whether the running time is linear,
polynomial or otherwise (as a function of input values);
and which computed values are polynomially
(or linearly) bounded. 
Suprisingly, the analysis itself takes \emph{polynomial time}.

Once a result of this kind is established, a program of further research arises
automatically: investigate the tradeoff between the strength of the core language 
(the abstraction) and the decidability of the properties of interest. 
A stronger core language can model more closely the concrete semantics of
real programs; in other words, it constitutes a finer abstraction. Make it too fine, and
decidability will be lost. It is interesting to find out how far we can venture
while maintaining decidability: what language features are
the real impediments? What extensions will increase the difficulty of the analysis?
Note that since we are dealing with decidable problems, we can classify their complexity.
Thus, in our approach, if  a polynomial algorithm for some problem is not found, we can
look for a hardness proof to justify it.

This paper makes a contribution to this program by analysing a mild extension to
the core language of~\cite{BJK08}. Specifically, we add resetting assignments (\verb/X:=0/). 
It is shown that the analysis problem becomes PSPACE-hard.
This is tight: a PSPACE algorithm is given.
The algorithm
evolved from the previous solution in an interesting way:
focus was shifted from proving a bound to disproving it, and
the algorithm works top-down rather than bottom-up.
 
\paragraph*{Related work.}
Prior to our work in~\cite{BJK08}, Niggl and Wunderlich~\cite{NW06} and later
Jones and Kristiansen~\cite{JK08} studied similar languages with similar methods---except
that their programs were ``too concrete'' and their analyses sound but incomplete.
While both works consider branch conditionals as ``uninterpreted,'' which technically
means the same as abstracting them the way we do here, the loops are considered 
as exact (in contrast to the non-deterministic loops used here, see the following
section).

An important feature of our core language, crucial in our analyses, is
being structured (which allows for a compositional analysis, or proof
rules, as in this paper). In contrast,
Albert et al.~\cite{AAGP:sas08} propose an ``intermediate language'' for
complexity analysis that is essentially unstructured---quite natural,
given that its original application was to Java bytecode. They emphasize
that this language ``can be formally defined independent of the
[source] programming language.'' This agrees with our approach, however
their language is too strong---it is, in fact, Turing-complete for
computations over integers---so its analysis is as undecidable as for any
concrete language.


\section{Problem Statement}
\label{sec:goals}

First, we define our core language, $L_r$.

\paragraph{Syntax} is described in Figure~\ref{fig-syntax} and should be self-explanatory.
In a command $\verb+loop+\; \X \; \verb+{C}+$,
variable $\X$
is not allowed to appear on the left-hand side of an assignment in the loop
body {\tt C}. There is no special syntax for a ``program.'' 

\begin{figure}[t]
$$ \renewcommand{\arraystretch}{1.3}
\begin{array}{rcl}
\verb+X+,\verb+Y+\in\mbox{Variable} &\;\; ::= \;\; & \X_1 \mid\X_2 \mid \X_3 \mid
 \ldots  \mid \X_n\\
\verb+e+\in\mbox{Expression} & ::= & \verb+X+ \mid \verb/X + Y/ \mid 
\verb+X * Y+\mid\verb+0+\\
\verb+C+\in\mbox{Command} & ::= & \verb+skip+ \mid \verb+X:=e+ 
                                \mid \verb+C+_1 \semi \verb+C+_2 
                                \mid \texttt{loop} \; \X  \; \texttt{\{C\}} \\
                         & \mid & \texttt{choose}\;  \C_1  \; \texttt{or} \; \C_2
 \end{array} \renewcommand{\arraystretch}{1.0}$$

\caption{Syntax of the core language $L_r$. Variables hold nonnegative
integers. \label{fig-syntax}}
\end{figure}

\paragraph{Data.}
It is most convenient to assume that the only type of data is
nonnegative integers.
More generality is possible but will not be treated in this presentation.

\paragraph{Command semantics.}
The semantics of the core language is intended for over-approximating a realistic
program's semantics. Therefore, the core language is nondeterministic.
The {\tt choose} command represents a nondeterministic choice, and can be used to abstract
any concrete conditional command by simply ignoring the condition.
The {\em loop command\/} 
$\verb+loop+\,\X_\ell\,\verb+{C}+$ repeats \pgt{C} a number of times
bounded by the value of $\X_\ell$. Thus, it 
may be used to model different kinds of loops (for-loops, while-loops)
as long as a bounding expression can be statically determined (possibly by
an auxiliary analysis such as~\cite{CS:01,PR:04}).

The use of bounded loops restricts the computable functions
to be primitive recursive, but this is still rich enough to make
the analysis problem challenging.

Formally, the semantics
associates with every command
\verb+C+ over variables $\X_1,\dots,\X_n$
a relation $\sempar{C} \subseteq \nats^n \times \nats^n$.
 In the expression $\vec x \sempar{C} \vec y$,
vector $\vec x$ (respectively $\vec y$) is the store before (after) the
execution of \verb+C+.

The semantics of {\tt skip}
 is the identity. The semantics of an \emph{assignment} leaves some
room for variation: either the precise value of the expression is assigned,
or a nonnegative integer bounded by that value. Because our 
analysis only involves monotone increasing functions,
this choice does not affect the results; but the second, non-deterministic definition increases
the coverage of the core language, since   
additional numeric operations can be modelled (over-approximated).
Finally, composite commands are described by the 
equations:
\begin{eqnarray*}
\lsem{\tt C}_1 {\tt ;C}_2\rsem &=&
  \sempar{C$_2$}\circ\sempar{C$_1$} \\
\lsem\verb+choose C+_1\verb+ or C+_2\rsem &=&
  \sempar{C$_1$}\cup\sempar{C$_2$} \\
\lsem\verb+loop+\; \X_\ell \; \verb+{C}+\rsem &=&
\{ (\vec{x},\vec{y}) \mid \exists i \le x_\ell :
\vec{x} \sempar{C}^i \vec{y} \}
\end{eqnarray*}
where $\sempar{C}^i$ represents $\sempar{C}\circ\cdots\circ\sempar{C}$
($i$ occurrences of $\sempar{C}$); and $\sempar{C}^0 = \sempar{skip}$.

For every command we also define its {\em step count\/} (informally
referred to as running time). The step count of an
atomic command is defined as 1. The step count of a loop command is the
sum of the step counts of the iterations taken. Because of the
nondeterminism in \pgt{if} and \pgt{loop} commands, the step count is also
a relation.  

\paragraph{Goals of the analysis.}  
The \emph{polynomial-bound analysis problem} is to find,
for any given command, which output variables
are  bounded by a polynomial in the input variables. This is the problem we will fix
our attention on, although some variants may also be of interest:
The \emph{linear-bound problem} identifies linearly-bounded output values instead.
The \emph{feasibility problem} finds whether all the values througout any computation
of the command are polynomially bounded. This problem differs from the 
polynomial-bound problem, because an exponential value may be computed and discarded.
But it can be solved with a small modification to the algorithm.
The \emph{polynomial (linear) step-count problem} is what its name implies, and can
be reduced to the the corresponding bound analysis problem by simple means
(see~\cite{BJK08}).

\paragraph{An example.} In the following program,
variables may grow exponentially (the reader is invited to check).
\begin{Verbatim}[codes={\catcode`$=3\catcode`_=8}]
loop X$_4$ {
   X$_3$ := X$_1$+X$_2$;
   choose { X$_1$ := X$_3$ } or { X$_2$ := X$_3$ };
 }
\end{Verbatim}
However, the following version is polynomially bounded:
\begin{Verbatim}[codes={\catcode`$=3\catcode`_=8}]
loop X$_4$ {
   X$_3$ := X$_1$+X$_2$;
   choose { X$_1$ := X$_3$ } or { X$_2$ := 0 };
 }
\end{Verbatim}


\section{Background: the Analysis of~\cite{BJK08}}

The language treated in~\cite{BJK08} differs from our $L_r$ only on
one essential point: the constant 0 is not included. 
Another difference is that expressions could be nested; but assignments
with nested expressions can be easily unfolded to unnested ones.
Let $\Lbjk$ denote the language obtained from $L_r$ by omitting the
constant 0. Then from~\cite{BJK08} we have

\bthm
The polynomial-bound analysis problem for $\Lbjk$ is in PTIME.
\ethm

The algorithm that does this evolved from ideas in
previous papers such as~\cite{NW06,JK08}.  These works, too,
considered a structured imperative language, and gave an algorithm
that evaluates any given command to a so-called
\emph{certificate}. The certificate is a finite structure
(basically, a matrix)
which summarizes the input-output dependences in the command: how each of
the output values depends on each of the inputs.
The analysis is compositional, which means that
a certificate for a composite command only depends on those of its parts.
In fact, it parallels the (compositional) definition of semantics,
and can be stated as an abstract interpretation~\cite{Cousot:ACM:96}.
The main novelty in the algorithm of~\cite{BJK08} was
a new kind of certificates, which encode sufficient information for
precise analysis of the chosen core language.

One of the open problems left by~\cite{BJK08} was analysis of extended
versions of the core language, and this paper presents a first step
forward. The extension considered is the capability to
reset a variable to zero.
One way to explain the difficulty caused by resets is as increased
context-sensitivity: for example, the command \verb/X:=Y*Z/ 
introduces a dependence of \pgt{X} on \pgt{Z},
but not if \pgt{Y} is fixed at zero.
We shall solve this problem by employing context-sensitive
analysis~\cite{Nielson-Nielson-Hankin}.
 When applied to the compositional (bottom-up) approach, this would mean
creating a set of certificates for a given command, each annotated with the
context to which it applies, and indicating the context that results.
This would create an explosion in the algorithm's complexity: the results
of analysing a command would be exponentially big. We want to avoid that
(though polynomial \emph{time} must be given up---witness the hardness
result in Section~\ref{sec:hardness}). This is why this paper presents not
just an extension, but a redesign of the analysis of $\Lbjk$.

The design of the new solution was motivated by an observation on the
correctness proofs of~\cite{BJK08}: the
soundness theorem (if the algorithm concludes that an output is
polynomially bounded, it really is)
was more difficult to prove than the completeness
theorem (if the algorithm concludes ``not polynomial," this is correct).
In retrospect, there is a simple reason: an output variable may be a
function of several input variables, and may depend on them differently
in different executions; the positive conclusion must account for all of
these dependences---it is a universal statement. A negative conclusion,
its negation, is existential: just find \emph{some} execution pattern in which the
output depends exponentially on \emph{some} input!
Noting this imbalance we will give, not an algorithm to certify
outputs as polynomial, but a \emph{proof system} to prove that an output can
be exponential\footnote{Which in this case is equivalent to not being polynomial.}.
Searching for a proof is naturally performed top-down, and it is in this way that
we can prove it to be in PSPACE. Thus, this design
is the converse of~\cite{BJK08} in a second sense: favouring the top-down
progress to bottom-up.

\section{A Proof System for (Lack of) Polynomial Bounds}
\label{sec-sop}

\subsection{Ingredients}

The basic ingredient in the proof system is called a \emph{dependence fact}.
In its simple (unary) form,
it indicates that an output variable depends, in a certain way, on some
input variable. The set of variable indices is denoted by $\inds$ with generic elements $i,j,k$
etc.

\bdfn
The set of \emph{dependence types} is
$\deptypes =  \{1,1^+,2,3\}$, with order $1 < 1^+ < 2 < 3$, and binary maximum operator
$\sqcup$.  We write $x\simeq 1$ for $x\in\{1,1^+\}$.
\edfn

Verbally, we may refer to these types as:

$1=$\emph{identity dependence},
$1^+=$\emph{additive dependence},
$2=$\emph{multiplicative dependence},
$3=$\emph{exponential dependence}.

\bdfn
The set of \emph{dependences} $\deps$ is the union of two sets: \\
(1) The set of \emph{unary dependences}, isomorphic to 
$\inds\times\deptypes\times\inds$. The notation for an element is
\ $\unarydep{i}{\delta}{j}$. \\
(2) The set of \emph{binary dependences}, isomorphic to 
$\inds\times\inds\times\inds\times\inds$, where the notation for an element is
\ $\binarydep{i}{j}{k}{\ell}$.
\edfn

Informally, a binary dependence represents a conjunction, namely the fact that
two unary dependences hold simultaneously.
 This is only used when the dependences in question are of types $1$ or
$1^+$, and when $i\ne j \lor k\ne \ell$
 (a similar mechanism was used in~\cite{BJK08}).  

\bdfn
A \emph{context} is a subset of $\inds$. A \emph{dependence judgement} is
$\C,P \vdash D,Q$ where $\C$ is a command, $P,Q$ are contexts and $D$ is a dependence.
\edfn

The pre-context $P$ specifies variables that are presumed to hold zero; the post-context
$Q$ specifies zeros guaranteed to come out.
 For an example, let $\C$ be the command
 $\verb+loop+\; \X_3 \; \verb+{X+_1\verb/:= X/_2\verb/+X/_3\verb+}+$.
We have $ \C, \emptyset  \vdash  \unarydep{2}{1}{2},   \emptyset$ ($\X_2$ is not
modified) and $ \C, \emptyset  \vdash  \unarydep{2}{1^+}{1},   \emptyset$ ($\X_1$ may be
set to $\X_2$ plus something else).
If $\X_3$ is initially zero, the loop does not execute. Therefore
$\C, \{ 3 \}  \vdash  \unarydep{2}{1^+}{1},  Q $
does \emph{not} hold for any $Q$.
However, $\C, \{ 3 \}  \vdash  \unarydep{1}{1}{1},  \{ 3 \}$
holds: $\X_1$ is not modified and $\X_3$ is guaranteed to remain zero.

 The inference rules for assignment commands, listed
 next, may further clarify the function of judgements.
 
\subsection{Inference rules for assignments}
 
 We list the {\tt skip} command among the assignments. It is, in fact, equivalent to
 $\verb+X+_1\verb/:= X/_1$.  Nonetheless, it is instructive to examine it first.
 
\be
\item (Unary rule for \verb+skip+) \par
\[\frac{ i\notin P } { \texttt{skip},\, P  \vdash \unarydep{i}{1}{i},\, P } \]

\item (Unary rule for $\X_l\verb+:=0+$) \par
\[\frac{ i\notin P,\ i\ne l } {
 \X_l\texttt{:=0},\, P  \vdash \unarydep{i}{1}{i},\, P \cup \{ l \}
}\]

\item (Unary rules for  $\verb+X+_l\verb/:=X/_r$) \par
 For any context $P$, let
$P_{\ell,r} = P \setminus \{l\} \cup \{l\mid \text{ if $r\in P$}\}$. 
\[\frac{ i\notin P,\  i\ne l}
{\X_l\texttt{:=X}_r,\, P  \vdash \unarydep{i}{1}{i},   P_{\ell,r}}
\qquad
\frac{r\notin P}
{\X_l\texttt{:=X}_r,\, P  \vdash \unarydep{r}{1}{l},   P_{\ell,r}}
\]

\item (Unary rules for  $\verb+X+_l\verb/:=X/_r\verb/*X/_s$) \par
\label{itm:rules-mult}
  For any context $P$, let
$P_{l,r,s} = P \setminus \{l\} \cup \{l\mid \text{ if $r\in P$ or $s\in P$}\}$.

\[\frac{i\notin P,\ i\ne l}
{ \X_l\texttt{:=X}_r\texttt{*X}_s,\, P  \vdash \unarydep{i}{1}{i},\, P_{l,r,s} }
\qquad
\frac{r,s\notin P,\ t\in \{r,s\}}
{\X_l\texttt{:=X}_r\texttt{*X}_s,\, P  \vdash \unarydep{t}{2}{l},\, P_{l,r,s}
}\] 

\item (Unary rules for  $\verb+X+_l\verb/:=X/_r\verb/+X/_s$, where $r\ne s$) \par
  For any context $P$, let
$P_{l,r,s} = P \setminus \{l\} \cup \{l\mid \text{ if $r,s\in P$}\}$.

\begin{tabular}{*{2}{p{0.4\textwidth}}}
\[\frac{i\notin P,\ i\ne l}
{\X_l\texttt{:=X}_r\texttt{+X}_s,\, P  \vdash \unarydep{i}{1}{i},\, P_{l,r,s} }
\]
&
\[\frac{r\notin P,\ s\in P}
{ \X_l\texttt{:=X}_r\texttt{+X}_s,\, P  \vdash \unarydep{r}{1}{l},\, P_{l,r,s} }
\] \\
\[\frac{r\in P,\ s\notin P}
{\X_l\texttt{:=X}_r\texttt{+X}_s,\, P  \vdash \unarydep{s}{1}{l},\, P_{l,r,s} }
\]
&
\[\frac{r,s\notin P}
{\X_l\texttt{:=X}_r\texttt{+X}_s,\, P  \vdash \unarydep{t}{1^+}{l},\, P_{l,r,s} 
  \text{ for $t\in\{r,s\}$.}
}\]
\end{tabular}

\item (Binary rules for assignments) \par
Let $\C$ be any of the above commands. If, for $i,i'\notin P$, and $j,j'\notin Q$,
where $i\ne i'$ or $j\ne j'$, we have
$\C,P  \,\vdash \,\unarydep{i}{r_1}{j},  Q$ and
$\C,P  \,\vdash \,\unarydep{i'}{r_2}{j'},  Q$, where $r_1,r_2\simeq 1$, then 
$\C,P  \,\vdash \,\binarydep{i}{i'}{j}{j'},  Q$.
\ee

\subsection{Inference rules for composite commands}

The composite commands are the choice, sequential composition and
the loop.

\paragraph*{Choice} is simplest, handled by the obvious rules:

\begin{equation}\mytag{C}
\frac{   {\tt C}_1, P \vdash D, Q }
{ {\tt choose\,C}_1 {\tt or\,C}_2, P \vdash  D, Q }
\qquad
\frac{   {\tt C}_2, P \vdash D, Q }
{ {\tt choose\,C}_1 {\tt or\,C}_2, P \vdash  D, Q }
\end{equation}

\paragraph{Sequential composition} requires an operator for abstract composition,
that is, composition of dependences.
 
 \bdfn
 The binary operation $\cdot$ is defined on $\deps$ by the following rules:
\[\renewcommand{\arraystretch}{2} 
\begin{array}{ccc}
   (\unarydep{i}{\alpha}{j}) \cdot (\unarydep{j}{\beta}{k}) &=& 
   \multicolumn{1}{l}{\longunarydep{i}{\alpha\sqcup\beta}{k}} \\  
   (\unarydep{i}{\alpha}{j}) \cdot (\binarydep{j}{j}{k}{k'}) &=& (\binarydep{i}{i}{k}{k'}),
\quad\text{provided $\alpha \simeq 1$} \\
    (\binarydep{i}{i'}{j}{j}) \cdot (\unarydep{j}{\alpha}{k}) &=& (\binarydep{i}{i'}{k}{k}),
\quad\text{provided $\alpha \simeq 1$} \\
   (\binarydep{i}{i'}{j}{j'}) \cdot (\binarydep{j}{j'}{k}{k'}) &=& 
\left\{\begin{array}{cl}
\binarydep{i}{i'}{k}{k'},  &  \text{ if $i\ne i'$ or $k\ne k'$} \\
\unarydep{i}{2}{k},  &  \text{if $i=i'$ and $k= k'$}
\end{array}\right.
 \end{array}
\]
 \edfn
We now have the rule
\begin{equation}\mytag{S}
\frac{   {\tt C}_1, P \vdash D_1, Q \quad {\tt C}_2, Q \vdash D_2, R }
{  {\tt C}_1 {\tt ;C}_2, P \vdash   {D_1\cdot D_2}, \, R }
\end{equation}
Naturally, the rule is only applicable if $D_1\cdot D_2$ is defined.

\paragraph*{The loop} involves the possibility of growth that depends on the number of 
iterations.  To handle this, we introduce a \emph{loop correction} operator (not unlike
the one in~\cite{BJK08}).

\bdfn \label{def:LC}
The loop correction operator $LC_\ell : \deps\to\deps$ is defined by
\begin{align*}
LC_\ell (\unarydep{i}{1^+}{i}) &= 
   \unarydep{\ell}{2}{i}   \\
LC_\ell (\unarydep{i}{2}{i}) &= 
   \unarydep{\ell}{3}{i}  
\end{align*}
\edfn
Explanation: in the first case, $\X_i$ has something added to it. Intuitively, if this happens
inside a loop, it
results in growth that is at least linear in the number of iterations.
In the second case, $\X_i$ is multiplied by something, which results in exponential growth.

There are three loop rules. The first covers the case that the body is not executed.
\begin{equation}\mytag{L0}
\frac{ \texttt{skip}, P  \vdash D, P } 
{{\tt loop\,X_\ell\{C\}}, P \vdash   D, P }
\end{equation}

The second describes the result of any number $m>0$ of iterations.
\begin{equation}\mytag{L1}
\frac{ {\tt C}, P_0  \vdash D_{1}, P_{1}\quad {\tt C}, P_1  \vdash D_{2}, P_{2}\quad\dots\quad
{\tt C}, P_{m-1}  \vdash D_{m}, P_{m}\qquad \ell\notin P_0 }
{{\tt loop\,X_\ell\{C\}}, P_0 \vdash   {D_1\cdot D_2\cdot\ldots\cdot D_m}, P_m }
\end{equation}

The third applies the LC operator.
\begin{equation}\mytag{L2}
\frac{ 
{\tt loop\,X_\ell\{C\}}, P_0  \vdash D_{1}, P_{1} \quad
{\tt loop\,X_\ell\{C\}}, P_1  \vdash D_{2}, P_{1} \quad
{\tt loop\,X_\ell\{C\}}, P_1  \vdash D_{3}, P_{3} \qquad \ell\notin P_0}
{{\tt loop\,X_\ell\{C\}}, P_0 \vdash   {D_1\cdot LC_\ell(D_2)\cdot D_3}, \, P_3}
\end{equation}

Note that as $LC_\ell$ is applied to $D_2$, we require that $D_2$ be a dependence
that can be iterated: this requires that the pre-context and post-context be
the same. An example may help to clarify this rule. Consider first the following loop {\tt L}:

\begin{Verbatim}[codes={\catcode`$=3\catcode`_=8}]
loop X$_4$ { 
    loop X$_3$ { X$_1$ := X$_2$ } ;
    X$_2$ := X$_1$ + X$_2$; 
}
\end{Verbatim}

It is easy to see that the outer loop's body, {\C}, consisting of the inner loop and the subsequent
assignment, may set $\X_2$ to twice its initial value. This is expressed by the
judgement $\emptyset, \C \vdash \unarydep{2}{2}{2}, \emptyset$
(the reader may want to construct the proof for this);
it therefore follows by (L1) that 
\begin{equation} \label{D1}
{\tt L}, \emptyset \vdash \unarydep{2}{2}{2}, \emptyset \,.
\end{equation}
It is also the case that 
\begin{equation} \label{D2}
{\tt L,} \emptyset \vdash \unarydep{2}{1}{1}, \emptyset 
\end{equation}
and by (L0),
\begin{equation} \label{D0}
{\tt L,} \emptyset \vdash \unarydep{4}{1}{4}, \emptyset \,.
\end{equation}
Since the pre-context and the post-context are the same in (\ref{D1}), this judgement describes
an \emph{iterative} dependence and by (L2) we obtain
\[
{\tt L}, \emptyset \vdash   { (\unarydep{4}{1}{4})\cdot LC_\ell(\unarydep{2}{2}{2})\cdot (\unarydep{2}{1}{1})}, \, \emptyset 
\]
that is
\[
{\tt L}, \emptyset \vdash   \unarydep{4}{3}{1}, \, \emptyset
\]
The result in $\X_1$ depends exponentially on the input $\X_4$.

Change now the loop as follows:

\begin{Verbatim}[codes={\catcode`$=3\catcode`_=8}]
loop X$_4$ { 
    loop X$_3$ { X$_1$ := X$_2$ } ;
    X$_2$ := X$_1$ + X$_2$; 
    X$_3$ := 0;
}
\end{Verbatim}
Now instead of (\ref{D1}), we have
\begin{equation} 
{\tt L}, \emptyset \vdash \unarydep{2}{2}{2}, \{3\}
\end{equation}
Where the post-context represents the resetting of $\X_3$.
This dependence is not iterative: the post-context differs from the pre-context. (L2) is not applicable as above and indeed,
the modified loop does not generate exponential growth any more.

Another subtle point with this rule is the role of the
``preamble'' $D_1$ that comes before the iterative part.
Since the $LC_\ell$ operator only expresses
dependence on the loop variable $\X_\ell$, and this
variable is not modified within the loop, the dependence $D_1$
can only be $\unarydep{\ell}{1}{\ell}$. Why is it included
then? The reason is that a certain, finite number of
initial iterations may set up a \emph{context} necessary for
the iterative part. Consider the following command.

\begin{Verbatim}[codes={\catcode`$=3\catcode`_=8}]
X$_2$ := 0;  loop X$_3$ { X$_1$ := X$_2$*X$_1$; X$_2$ := X$_1$ } 
\end{Verbatim}

The loop command, {\tt L}, 
satisfies 
${\tt L},\emptyset \vdash \unarydep{1}{2}{1}, \emptyset$
which leads to exponential dependence of  $\X_1$ (and consequently
$\X_2$) on $\X_3$. But 
the loop is entered in the context $\{2\}$ (note the
preceding reset), and $ {\tt L},\{2\} \vdash
\unarydep{1}{2}{1}, \{2\}$ does not hold.
It is therefore
crucial to note that an initial iteration can modify $\X_2$.
This appears formally as:
\begin{equation}
\frac{ 
{\tt L}, \{2\}  \vdash \unarydep{3}{1}{3}, \emptyset \quad
{\tt L}, \emptyset  \vdash \unarydep{1}{2}{1}, \emptyset \quad
{\tt L}, \emptyset  \vdash \unarydep{1}{1}{2}, \emptyset \qquad 3\notin \{2\}}
{\displaystyle \frac{
{\tt L}, \{2\} \vdash  (\unarydep{3}{1}{3})\cdot LC_\ell(\unarydep{1}{2}{1})\cdot (\unarydep{1}{1}{2}), \, \emptyset}
{{\tt L}, \{2\} \vdash  \unarydep{3}{3}{2} ,\, \emptyset}
}
\end{equation}

\subsection{Correctness theorem}

\bthm \label{thm:poly-correct}
Let $\C$ be a command of $L_r$ (interpreted over the non-negative integers).
If $\C,\{ \} \vdash \unarydep{i}{3}{j}, Q$ for some $i$ and $Q$, then the values 
that $y_j$ can take, given $\vec x \sempar{C} \vec y$, cannot be bounded polynomially
in $\vec x$. If no such judgement can be proved,  a polynomial bound exists.
\ethm

While the full proof will not be given here, here is a very brief overview.
The first direction (soundness of the conclusion of exponential growth) is proved
in the natural way: after defining carefully the semantics of judgments, one proceeds
to prove that each inference rule is sound. This requires some care in the case of loops,
while the rest is straight-forward.

The second direction (soundness of the conclusion of polynomial boundedness)
requires more effort. It is helpful to view the analysis as an abstract interpretation
where the abstract semantics of a command is the set of judgements that can be
proved about it.  We prove that when this set does not contain a Type~3 dependence for
$\X_j$, its final value admits a polynomial bound. Clearly, this set can be exponentially big,
but without contexts it is of polynomial size. Thus, the result of~\cite{BJK08} can be reconstructed
as a special case.

\section{Checking for Linear Bounds}
\label{sec-lin}

In Section~\ref{sec:goals} a few variants of our problem were described. Here we give
the details of solving one of them: the linear-bound problem.
This problem differs from the polynomial-bound problem only in the 
definition of ``good" and ``bad" growth: now every non-linear dependence is ``bad"
and will be labeled by a 3.
Thus, 2's only describe linear bounds.

For example,
 the command 
$\X_1\pass\X_1 \texttt{+} 2\texttt{*}\X_2$ (with empty pre-context) satisfies
  $\dep{\mbox{${\tt X}_2$}}{1}{\mbox{${\tt X}_2$}}$,
\ $\dep{\mbox{${\tt X}_2$}}{2}{\mbox{${\tt X}_1$}}$,
\ $\dep{\mbox{${\tt X}_1$}}{1^+}{\mbox{${\tt X}_1$}}$;
while $\X_1\pass\X_1 \texttt{+} \X_2\texttt{*}\X_2$ yields
$\dep{\mbox{${\tt X}_2$}}{3}{\mbox{${\tt X}_1$}}$.

Judgements $\vdash_{lin}$ for linear bounds are proved just like their counterparts
for polynomial bounds (judgments $\vdash$) except as follows.
\begin{enumerate}
\item
The second unary rule for multiplication (Page~\pageref{itm:rules-mult}) is changed into
\[\frac{r,s\notin P,\ t\in \{r,s\}}
{\begin{array}{llrll}
\X_l\texttt{:=X}_r\texttt{*X}_s, P  &\vdash &\unarydep{t}{3}{l},  & P_{l,r,s}
  &\text{for $t\in\{r,s\}$.}
\end{array}}\]
\item
In the definition of the loop correction operator (Page~\pageref{def:LC})
one case is modified:
\[ LC_\ell (\unarydep{i}{1^+}{i}) = 
   \unarydep{\ell}{3}{i} \,.
\]
\end{enumerate}

\bthm \label{thm:lin-correct}
Let $\C$ be a command of $L_r$ (interpreted over the non-negative integers).
If $\C,\{ \} \vdash_{lin} \unarydep{i}{3}{j}, Q$ for some $i$ and $Q$, then the values 
that $y_j$ can take, given $\vec x \sempar{C} \vec y$, cannot be bounded linearly
in $\vec x$. If no such judgement can be proved,  a linear bound exists.
\ethm

\section{Complexity}

Here we establish the complexity class of the polynomial-bound (or linear-bound)
analysis problem for $L_r$.
Note that the complexity is in terms of the size of the analysed program.

\subsection{Inclusion in PSPACE}

Savitch's theorem makes it easy to prove inclusion in PSPACE by giving a polynomial-space
\emph{non-deterministic} algorithm, which is quite easy to do here.

Consider a recursive algorithm that tries to prove $\C,\{ \} \vdash \unarydep{i}{3}{j}, Q$
with $i$ and $Q$ non-deterministically chosen (and $j$ too, if we do not query a particular
output). The algorithm attempts to 
develop a proof tree non-deterministically, in a depth-first manner.
To ensure polynomial space complexity, it is necessary to bound the depth of the
proof tree, and the amount of memory needed to process each node.
The latter is easily seen to be
polynomial (this is the case even with rule L1, despite the fact that $m$ 
is not polynomially bounded, since the rule can be applied without
keeping the whole list of steps in memory).
As for the depth, for all commands except the loop, the subgoals of a proof node for the command
relate to its sub-commands; thus the proof tree would have precisely the depth of the
syntax tree, were it not for the loop rule L2, which recursively uses judgements on the
same loop. However, the following
lemma (stated without proof) solves the problem.

\blem
Let $\C$ be a loop command. If $\C,P\vdash D,Q$ can be proved using rule L2,
then it can be proved using L2 where the premises are only proved using L0 and L1.
\elem

\subsection{Hardness}
\label{sec:hardness}

We prove PSPACE-hardness of the linear-bound problem. The proof can be tweaked
to apply to polynomial bounds as well.
It is a reduction from a binary-alphabet
 version of the well-known PSPACE-complete problem,
\emph{NFA Universality}. The problem is defined as follows.

\begin{quotation}
INSTANCE: A nondeterministic finite automaton $A=(Q,\Sigma,\delta,q_0,F)$,
where $\Sigma=\{0,1\}$, $\delta\subseteq Q\times\Sigma\times Q$ is the set of
transitions, $q_0$ the initial state and $F$ the (non-empty) set of accepting states.
\end{quotation}

\begin{quotation}
QUESTION: Is $L(A)=\Sigma^*$?
\end{quotation}

Let $A$ be given, with $Q=\{1,\dots,n\}$.
We construct a program with 
variables $\X_1,\dots,\X_n,\X'_1,\dots,\X'_n,\Y,\Z$.  For convenience we use compound
expressions (but these can be taken to be syntactic sugar).

For each $q\in Q$ and $a\in\Sigma$, let $\C_{q,a}$ be the command
\[ \X'_q \verb/:= Z*X/_{i_1}\!\verb/*X/_{i_2} \verb/*/\ldots\verb/*X/_{i_p} \]
where $\{i_1,\dots,i_p\}$ are all the predecessors of $q$ under $a$-transitions.
The multiplication by \pgt{Z} solves the difficulty with states having no predecessors.

For each $a\in \Sigma$, let $\C_{a}$ be
\[  \C_{1,a} {\tt ;\ } \C_{2,a} {\tt ;\ } \dots \C_{n,a} {\tt ;\ }
   \X_1\verb/:=/\X'_1{\tt ;\ }\dots {\tt ;\ } \X_n\verb/:=/\X'_n
\]
and let \pgt{Fin} be the command 
\[ \verb/Z := Z*X/_{i_1}\!\verb/*X/_{i_2}\cdots \verb/*X/_{i_{|F|}} \]
where $F = \{i_1,\dots,i_{|F|}\}$.

Let \pgt{P} be the program:
\[ \X_{q_0} \verb/:=0; loop Y {choose C/_0 {\tt \ or}\ \C_1\verb/}; Fin/ \]


Suppose that the loop performs
$k$ iterations, where the $i$th
iteration chose $\C_{a_j}$. At its end, variable $\X_q$ will be 0 if and only if state $q$ is
reached on the word $w = a_1\dots a_k$.   In \pgt{Fin}, $\Z$ will be set
to 0 if and only if at least one accepting state is reached, that is, if $A$ accepts
$w$. If it does not, $\Z$ will be set to 
a non-linear function of the inputs (and this can then happen for all inputs satisfying
$\Y \ge k$).
We conclude that $\Z$ is linearly bounded if and only if $A$ is universal.

Note that by this proof, PSPACE-hardness holds for $L_r$ without the addition
operation, since it is not used in the construction. Alternatively, since multiplication
can be simulated by a loop of additions, we can also conclude hardness for $L_r$
without multiplication. The choice command can also be eliminated, since it can be
simulated by the non-deterministic loop.  We obtain

\bthm
The linear-bound analysis problem is PSPACE-hard for the following subset of $L_r$:
\[\begin{array}{clcl}
\verb+X+,\verb+Y+&\in\mbox{Variable} &\;\; ::= \;\; & \X_1 \mid\X_2 \mid \X_3 \mid
 \ldots  \mid \X_n\\
\verb+e+&\in\mbox{Expression} & ::= & \verb+X+ \mid \verb/X + Y/ \mid {\tt 0}\\
\verb+C+&\in\mbox{Command} & ::= &  \verb+X:=e+ 
                                \mid \verb+C+_1 \verb+;C+_2 
                                \mid \verb+loop X {C}+
\end{array}\]
\ethm
The same holds for the other problems enumerated in Section~\ref{sec:goals}.

\section{Can we take it further?}
\label{sec:extensions}

There are many directions in which this work in particular, and the approach in a looser sense,
can be taken further.  The most immediate and obvious question to ask is how else can
$\Lbjk$ be extended while keeping our problems solvable. 
In other words, we would like to understand what programming-language features lend 
themselves to growth-rate analysis and which make trouble.

From our work in~\cite{BJK08}, one of the clear insights was that key to our success was
the property of \emph{monotonicity}. Firstly, of the functions computed by arithmetic
operators that we admitted---and indeed, if we extend
$\Lbjk$ with subtraction of integers (assuming exact arithmetics),
 we obtain a language where none of our
problems is decidable.

Monotonicity is also necessary in the semantics of commands. While for the 
\verb/choose/ command and sequential composition, a suitable notion of
monotonicity can be easily established,
 the loop command is more difficult. What we found was that
\emph{exact} loop semantics, where the number of iterations specified by the loop 
header is always completed, destroys monotonicity:  a second iteration may
erase the result of the first, so that increasing the iteration count does not increase the
output. Our non-deterministic loop semantics restores monotonicity, since increasing the
loop variable only makes more outcomes possible. That is the key to success, and happily,
it may also be useful (the kind of loops with non-monotone behaviour are probably not
a natural programming style, but loops that allow premature termination are).
Moreover, exact loops do defy analysis---and not much is required besides! 
The following tiny language has an undecidable linear-bound analysis problem, given that the
loop semantics is exact~\cite{BK:CiE09}:

\[\begin{array}{rcl}
\verb+X+,\verb+Y+\in\mbox{Variable} &\;\; ::= \;\; & \X_1 \mid\X_2 \mid \X_3 \mid
 \ldots  \mid \X_n\\
\verb+C+\in\mbox{Command} & ::= &  \verb+X:=Y+ \mid \verb/X:=Y+Z/
                                \mid \verb+C+_1 \semi \verb+C+_2 
                                \mid \texttt{loop} \; \X  \; \texttt{\{C\}}
\end{array}\]

From this point of view, we may explain the difficultly of including resetting assignments in the
analysis of
\cite{BJK08} by observing that they introduce a pinch of non-monotonicity. But not too
much: the introduction of contexts solves this problem. 
If you know where the zeros are, the rest is monotone.
Let us consider a few other possible extensions.

\subsection*{$\max$ and $\min$ operators}
The command \verb/X := max(Y,Z)/, with the natural semantics (over non-negative integers),
can be added to $L_r$ completely for free. It is, for the purpose of our analysis,
 equivalent to \verb/choose X:=Y or X:=Z/.

Not so for \verb/X := min(Y,Z)/. It is non-monotone, and adding it to $\Lbjk$ makes
the analysis PSPACE-hard, by essentially the reduction of Section~\ref{sec:hardness}
(minimum can play the
role of multiplication by zero).  Conjecture: this problem, too, is PSPACE-complete.

\subsection*{The constant 1}
Allowing the constant 1 in is a significant extension. 
Note that once you have 1, you can generate other positive constants. You can also 
compute fractions, in a sense: for example,
The following loop repeatedly exchanges variables,
so that the value increases every other iteration:
$$
\X_1 \pass \Z\semi\; 
\X_2 \pass \Z\semi\; 
\texttt{loop} \; \Y \;  \texttt{\{}
\, \Z \pass \X_1\semi\;  \X_1 \pass \X_2+1\semi\;
 \X_2 \pass \verb/Z }/
$$
Thus the effect of this loop is described by
the pseudo-command \verb/Z := Z+/$\frac{1}{2}\Y$.


It remains an open problem whether the analysis problems for this language are decidable.

\subsection*{Space bounds}
We mentioned the size of integers and running time. What about space bounds?
If we are interested in the number of bits taken 
by an integer-manipulating program, then bounding the numbers polynomially
is equivalent to bounding the space linearly. For another setting,
assume that the program allocates arrays by specifying their size,
and that the arrays can be left out of the abstraction (because there
is no data flow from the contents of the arrays to
the variables used as allocation sizes and loop indices; this is true for many programs).
Then one can test for polynomial space, essentially by including a
variable that sums all allocations.
A greater challenge, left for future work,
is the analysis of programs that can allocate as well as de-allocate
memory blocks; such programs are likely to use much less space then the
sum of all allocations (but note that our analysis cannot handle
subtraction!).

\subsection*{Other challenges}

\begin{itemize}
\item
Extending the types of growth-rates that can be decided beyond the current
selection of linear and polynomial.
In light of the results in~\cite{KN03,KN04}, it seems plausible that the method
could be extended to classify the
Grzegorczyk-degree of the growth rate.
\item
Moving to different language styles, such as programs with recursion, programs with high-order functions,
or unstructured (flowchart) programs.
\end{itemize}

\section{Conclusion}

This paper presents progress in a project of investigating program abstractions
with decidable growth-rate analysis problems, and establishing the computational
complexity of such problems.
In particular, we have added a reset-to-zero command to the language of~\cite{BJK08}.
We have suggested a solution based on inference rules that are targeted
at proving non-polynomial growth, instead of the previous approach of computing
certificates to polynomiality. This idea may be interesting in its own right; note that
we can freely exchange a problem with its complement precisely because we are designing 
a precise decision procedure!
This also allows us to classify the complexity of the analysis problem.
In the case of resets, we have
shown that the analysis problem rises from PTIME to PSPACE-complete.
There are other possible extensions whose complexity (even decidability) are not known.
An intriguing theoretical question is: are there any (reasonably natural) ``core languages"
for which the analysis is decidable, yet has a complexity above PSPACE-complete?

\subsection*{Acknowledgement}
The author is indebted to Lars Kristiansen for some helpful comments on the manuscript.

\bibliographystyle{eptcs}
\bibliography{icc}

\end{document}